\documentstyle[prd,aps,eqsecnum,preprint,tighten,floats]{revtex}
\begin{document}
\draft

\preprint{\vbox{\hfill pin97a.tex}
          \vbox{\vskip0.5in}
          }

\title{Wilson Loop on a Light-Cone Cylinder}

\author{Stephen S. Pinsky}
\address{Department of Physics, The Ohio State University, Columbus,
OH 43210}

\date{\today}

\maketitle

\begin{abstract} 
QCD without matter and quantized on a
light-cone spatial cylinder is considered. For the gauge group SU(N) the
theory has $N-1$ quantum mechanical degrees of freedom, which describe the
color fux that circulates around the the spatial cylinder. In $1+1$
dimensions this problem can be solved analytically. I use the solution
for SU(2) to compute the Wilson loop phase on the surface of the cylinder
and find that it is equal to $g^2 area/4$. This result is different from
the well known result for flat space. I argue that for SU(N) the Wilson
loop phase for a contour on a light-cone spatial cylinder is $g^2(area)
(N-1)/4$. The underlying reason for this result is that only the
$N-1$ dimensional Cartan subgroup of $SU(N)$ is dynamical in this
problem.
\end{abstract}
\pacs{ }

\newpage

%%%%%%%%%%%%%%%%%%%%%%%%%%%%%%%%%%%%%%%%%%%%%%
\section{Introduction}
%%%%%%%%%%%%%%%%%%%%%%%%%%%%%%%%%%%%%%%%%%%%%%

The expectation value of the Wilson loop is an important quantity in
gauge theories. This phase is a gauge invariant quantity that provides
information about the long range behavior of a theory, however it is beyond
the reach of weak coupling perturbation theory calculations. On the other
hand for non-perturbative approaches the Wilson loop is
an important object to consider. While for QCD in 3+1 dimensions the
calculation of the Wilson loop is quite difficult, in 1+1 dimension
it is much more tractable and in some problems can be calculated exactly.

The problem of pure glue $QCD$ in 1+1 dimensions in the gauge SU(N) with
periodic boundary conditions can be solved solved exactly
\cite{cyl,kap94} since it has only $N-1$ degrees of freedom which are
independent of space and the problem is therefore a quantum mechanical
rather than a true field theory problem.  Nevertheless the problem is very
interesting from a number of points of view. First the degrees of freedom
are simply color flux loops that circulate around the entire spatial
cylinder and as such they rely on the fact that the problem is formulated
on a cylindrical topology.  This problem is particularly interesting to
people studing light-cone field theory since it is the only known gauge
theory where the Hamiltonian takes exactly the same functional form in
both the light-cone and equal-time formulations. 

We will briefly review the formulation and solution of this problem, here
using the light-cone gauge and light-cone quantization in the gauge
$SU(2)$. We will solve for the wavefunctions and the energy eigenvalue
of the problem. We will solve the equations of motion for the vector
potential and use the solution to calculate the path integral of the vector
potential around a closed loop on the surface of the cylinder which makes
up space time in this problem. We then calculate the Wilson loop by
taking the vacuum expectation value of this loop calculation. 

There is an exact general result for the Wilson loop expection value in 1+1
dimensional QCD in the absence of matter. The result we find here by direct
calculation for $SU(2)$ on a cylinder does not agree with this result.
This is perhaps not surprising because of the special topology of the
space we consider. Based on our result for $SU(2)$ we sugest
a general result for the value of the Wilson loop phase for $SU(N)$ on a
cylinder. Our conjecture agrees with the general result in the large $N$
limit as would be suggested by the work of Gross\cite{gro93}.

\section{Gauge Fixing}

The Lagrangian density  for SU(2) non-Abelian gauge
theory in 1+1 dimensions is, 
\begin{equation}
{\cal L} = {1\over 2}\, {\rm Tr}\, (F_{\mu \nu} F^{\mu \nu})  
\end{equation}
where   
$F_{\mu \nu} = \partial _{\nu} A_{\nu} - \partial_{\nu} A_{\mu} -g[A_{\mu}, 
A_{\nu}]$. 
We consider the theory on a finite interval, $x^-$ from $-L$ to $L$, and we
impose periodic boundary conditions  on all gauge potentials $A_\mu$.  
 
We now show that the light-cone gauge $A^+=0$ which is the one that one
normal would prefere to use for light-cone quantization cannot be reached.  A
gauge transformation $U$ bringing a gauge potential $B^\mu$, itself in some
arbitrary gauge configuration, to some other gauge configuration
$A^\mu$ is 
\begin{equation}
gA^{\mu} = \partial_{\mu} U U^{-1} +g U B^{\mu} U^{-1}\;.
\label{gaugetrans} 
\end{equation}
Here $g$ is the coupling constant and $U$ is an element of the Lie algebra
of SU(2). Clearly $U$ given by  
\begin{equation}
U = P \exp{[-g \int_{-L}^{x^-} {dy^- B^+(y^-)}] } 
\label{naiveU}
\end{equation}
will bring us to the gauge $A{^+} = 0$.
  
We appear to have been successful in getting the light-cone gauge. 
However, the element $U$ through which we wish to achieve the gauge
condition must  satisfy $Z_2$ periodic boundary condition, $U(x^-) = (\pm)
U(x^-+2L)$. This is so, because gauge fixing is usually done with trivial
elements of the gauge group --- namely transformations generated by the
Gauss law operator via the classical brackets or corresponding quantum
commutators. However for this to be carried through, one needs to be
able to discard surface terms. With nonvanishing boundary conditions 
this can only be realized by $Z_2$ periodic elements $U$.  Clearly
Eq.(\ref{naiveU}) does not satisfy these boundary conditions. So in fact
the attempt has failed.

With  a modification of Eq.(\ref{naiveU}),
\begin{equation}
U(x) = e^{gx^-V^+} P e^{-g\int_{-L}^{x^-} dy^- B^+(y^-)}  
\;.
\end{equation}
where $V^+$ is the the integral $B^+$ over space normalized by the
length of the spatial cylinder, sometimes called the "zero mode",  this
is an allowed gauge transformation.   However it does not completely
bring us to the light-cone gauge.  We find instead 
\begin{equation}
A^+  = V^+ \;.
\end{equation} 
In other words, we cannot eliminate the zero mode of the gauge
potential. The reason is evident: it is {\it invariant} under periodic
gauge transformations. But of course we can always perform a rotation in
color space.   In line with other authors \cite{Fran81}, we choose this
so that $V^+ = v(x^+) \tau_3$ is  the only non-zero element, since in our
representation only $\tau^3$ is diagonal. 

In addition, we can impose the subsidiary gauge condition that the zero
mode of $A^-_3$ is zero. This would appear to have enabled complete
fixing of the gauge.  This is still not so. Gauge transformations
\begin{equation} 
G = \exp\{i x^- ({{n\pi} \over {L}}) {\tau}^3\}
\label{GribU}
\end{equation}
generate shifts, according to Eq.(\ref{gaugetrans}), in the zero mode
component 
\begin{equation}
v(x^+) \rightarrow v(x^+) + {{n\pi}\over{gL}}
\;.
\end{equation}
All of these possibilities, labelled by the integer $n$, of course still
satisfy $\partial_- A^+=0$, but as one sees $n=0$ should not really be
included. One notes that the
transformation is $x^-$-dependent and {\it $Z_2$ periodic}. It is thus a
simple example of a Gribov copy \cite{Grib78} in 1+1 dimensions.    We
follow the conventional procedure by demanding
\begin{equation}
v(x^+) \neq {n \pi \over gL}\;, \quad  n= \pm1, \pm2, \ldots 
\;.
\end{equation}
This eliminates singularity points at the Gribov `horizons' which
in turn correspond to a vanishing Faddeev-Popov determinant \cite{vBa92}.

The equations of motion for the theory are 
\begin{equation}
[D^{\mu}, F_{\mu \nu}] = \partial^{\mu} F_{\mu \nu} -g [A^{\mu},
 F_{\mu \nu}] = 0
\;.
\end{equation} 
For our purposes it is convenient to break this equation up into  color
components
$A{^{\mu} _a}$.  Color will always be the lower index.  Rather than the
three color fields $A{^{\mu} _1} , A{^{\mu} _2} ~{\rm and}~ A{^{\mu}
_3}$ we will use chiral notation with 
$ A{^{\mu} _+} = A{^{\mu} _1}  + iA{^{\mu} _2}$ and 
$A{^{\mu} _-} = A{^{\mu} _1} -iA{^{\mu} _2}$.   With the above gauge
conditions    the
$\nu=+$ equations are
\begin{eqnarray} 
(i\partial^+)^2{A^-_3} & = &  0, \\ 
(i\partial^+ + g v(x^+))^2{A^-_-} & = & 0
\label{3Gauss}
\end{eqnarray}   
These equation are of course easily solvable. The solution for $A^-_3$ 
is zero up to a constant which is the zero mode. Earlier we used our
gauge freedom to set this zero mode to zero. The operator in the
equation for $A^-_-$ is in fact not singluar in a particular Gribov
region and is therefore invertible giving  $A^-_-=0$. 

The only remaining equation of motion that is not totally trival is the
equation for $v$
\begin{equation}
\partial^2_+ v(x^+) = 0
\end{equation}
The solution is of course 
\begin{equation}
v(x^+) = { g^2 \over 2\pi} \Pi_zx^+ +v(0)
\end{equation}
and where $\Pi_z$ is the canonical momentum defined below. 

The Hamiltonian for this quantum mechanics problem is easily obtained
from the above Lagrangian and we find,
\begin{equation}
P^-= L\partial^2_+v(x^+)
\end{equation}
This leads to a set of properly normalized conjugate variables,
\begin{equation}
z={gLv \over \pi} \quad \Pi_z={2 \pi\over g} \partial_+v
\end{equation}
which satisfy thecanonical commution relation $[z,\Pi]=i$ in the
fundamental modular domain $ -1 < z < 0$. The Schr\"odinger equation is
straightforward to solve and wavefunction and energy eigenvalues are
\begin{equation}
\psi_n(z) = C_n sin(n\pi z) \quad E_n = {g^2L(n^2-1) \over 4}
\end{equation}
where we have renormalized the ground state ($n=1$) energy to zero and
$C_n$ are normalization constants. The wavefuntion must vanish at $z= 0$
and $-1$ the Gribov horizions\cite{kap94}. 

%%%%%%%%%%%%%%%%%%%%%%%%%%%%%%%%%%%%%%%%%%%%%%%%%%%%%%%%%
\section{Wilson Loop}
%%%%%%%%%%%%%%%%%%%%%%%%%%%%%%%%%%%%%%%%%%%%%%%%%%%%%%%%%

The vacuum expection value of the Wilson loop provides information about
the large distance behavior of a theory which is not accessable to  
pertubation theory calculation. The well established lore
associated with the Wison loop is that if the phase goes like the area of the
enclosed contour the theory is confining, whereas if the phase goes
like the perimeter of enclosed contour the theory is not confining.  

In 1+1 dimensions the general result for $SU(N)$ QCD without matter is
\cite{mig76},
\begin{equation}
W \propto e^{i g^2 {(N^2-1) \over 4N} {\cal A}},
\label{flatewl}
\end{equation}
where ${\cal A}$ is the area enclosed by the Wilson loop.
The Wilson loop for the problem we are considering here can be written  as the
vacuum expection value of the WIlson loop phase factor, 
\begin{equation}
W= \langle \psi_1 | Tr P e^{ig \oint A \cdot dx} | \psi_1\rangle.
\end{equation}
The vacuum expection value here takes the form of the expectation value
with the ground state wavefunction.  The contour that we will chose for the
path integral consists of straight lines connecting the following points in
$(x^-,x^+)$  space on the surface of the light-cone space-time cylinder.
\[
(0,0) \rightarrow (l,0) \rightarrow (l,t) \rightarrow (0,t) \rightarrow
(0,0)
\]
The only component of $A^\mu$ that is non-zero is $A^+=v(x^+)\tau_3$;
therefore the contour intergal yields $(v(0)-v(t))l\tau_3$. Now using
the solution of the equation of motion for $v(x^+)$ we find for the
contour integral,
\begin{equation}
ig\oint A\cdot dx =-i{g^2 \ 2\pi}{\cal A}\Pi_z \tau_3
\end{equation}
where ${\cal A}$ is the area of the enclosed contour. This leads to the
following expression for the Wilson loop W,
\begin{equation}
W=Tr\int_{-1}^0 dz sin(\pi z) ( cos(\theta) + i 
\tau_3 sin(\theta))sin(\pi z)
\end{equation}
where
\[
\theta = i{g^2 {\cal A} \over 4 \pi} {d \over dz}
\]
The momentum  operator acts on the ground state wave function to the right.
The expansion of the $sin(\theta)$ gives an odd number of derivatives leaving
an integral of $sin(\pi z) cos(\pi z)$ which vanishes when integrated over
the Gribov region $ -1 < z < 0$. This leaves only the
$cos(\theta)$ function and after some algebra I find
\begin{equation}
W \propto cos({g^2 {\cal A}\over 4})
\end{equation}
The general result Eqn(\ref{flatewl}) for the Wilson loop when
evaluated for $N=2$ gives ${g^2 3 {\cal A } \over 8}$ . 

%%%%%%%%%%%%%%%%%%%%%%%%%%%%%%%%%%%%%%%%%%%% 
\section{Discussion}
%%%%%%%%%%%%%%%%%%%%%%%%%%%%%%%%%%%%%%%%%%%%

Let us first summarize the essential points.  I analyzed pure glue
non-Abelian gauge theory in a  compact spatial volume with periodic
boundary conditions on the gauge potentials. Working in the light-cone
Hamiltonian approach, I demonstrated how one carefully fixes the gauge.
The quantum field theory problem then reduces to a quantum mechanical
problem which can be solved exactly. Given this exact non-perturbative
result for the vacuum state and vector potential it becomes a straightforward
calculation to evaluate  the Wilson loop and the result for the gauge group
$SU(2)$, $g^2 {\cal A}\over 4$ does not agree with the general result.

How can we understand these different results? The natural explanation seems
to be that on the cylinder the gauge field only has support on the
abelian Cartan subalgebra whereas the general result gets contributions from
all color components. We can speculate about the extension of this
calculation to SU(N) where the vector potential  only has support only 
$N-1$ dimensional abelian Cartan sub-algerbra. Since the contributions are
abelian we expect the phases to simply add for each additional field
component and therefore the Wilson loop phase should be 
\begin{equation}
W \propto e^{ig^2 (N-1){\cal A}\over 4}.
\label{cylwl}
\end{equation}
In a rather different context Gross \cite{gro93} has identified the
topological expansion of the space on which the  Wilson loop is
calculated with the $1/N$ expansion of the result.  This then allows us to
connect our calculation on a cylindrical topology with the large $N$
expansion of the general result Eqn(\ref{flatewl}). We see that in
the large $N$ limit Eqn(\ref{flatewl}) and Eqn(\ref{cylwl}) agree. Thus a
possible intepretation of this calculation might be the explicit
verification of the Gross \cite{gro93} result.

%%%%%%%%%%%%%%%%%%%%%%%%%%%%%%%%%%
\section*{Acknowledgments}
%%%%%%%%%%%%%%%%%%%%%%%%%%%%%%%
The author would like to acknowledge D. Robertson as a collaborator on
most of this work. This work is supported by a grant from the US Department
of Energy. 

%%%%%%%%%%%%%%%%%%%%%%%%%%%%%%%


\begin{references}
\bibitem{cyl}N. Manton, Ann. Phys. (NY) {\bf 159}, 220
(1985).  {}~J.E.~Hetrick,~Y.~Hosotani, Phys. Rev. D{\bf 38}, 2621 (1988);
J.E.~Hetrick,~Y.~Hosotani, PLB {\bf 30} (1989) 88,
F. Palumbo, Phys. Lett. B{\bf 243}, 109 (1990); 
S.G. Rajeev PLB {\bf 212} (1988) 203;
E. Langmann, G.W. Semenoff, Phys. Lett. B{\bf 303}, 303 (1993).
\bibitem{kap94} A. C. Kalloniatis, H.C. Pauli and S.S. Pinsky Phys. Rev.
{\bf D50}, 6633 (1994).
\bibitem{het93} J.E. Hetrick, Nucl. Phys. B (Proc. Suppl.) {\bf 30} 228
(1993), J.E. Hetrick UvA-ITFA 93-15 (hep-th/ 9305020).
\bibitem{mig76} A. Migdal, Sov. Phys. JETP {\bf 42} (1976) 413.
\bibitem{gro93} D. Gross, Nucl. Phys. {\bf 4 400} (1993) 161
\bibitem{PW}S.~J.~Brodsky, G. McCartor, H. C. Pauli, S. S. Pinsky,
Particle World  {\bf 3}, 109 (1993).
\bibitem{Lues83} M. L\"uscher, Nucl.
Phys. B{\bf 219}, 233 (1983); M. L\"uscher, G. M\"unster, Nucl. Phys. B{\bf
232}, 445 (1984). 
\bibitem{vBa92} P. van Baal, Nucl. Phys. B{\bf 369}, 259 (1992) and
references therein. 
\bibitem{Grib78} V.N. Gribov, Nucl. Phys. B{\bf 139}, 1 (1978);
H. Yabuki, Phys. Lett. B{\bf 231}, 271 (1989).
\bibitem{Fran81} V.A. Franke, Y.V. Novozhilov, E.V. Prokhvatilov,
Lett. Math. Phys. {\bf 5}, 437 (1981);~F.~Lenz, H.W.L. Naus, M. Thies,
`QCD in the Axial Gauge Representation', Erlangen preprint. To appear in
Ann. Phys.(NY) (1994).   
\end{references}
\end{document}